# Sequential File Programming Patterns and Performance with .NET

Peter Kukol

Jim Gray

Microsoft Research
Microsoft Corporation
One Microsoft Way
Redmond, WA  98052

## Table of Contents

**Abstract**: Programming patterns for sequential file access in the .NET Framework are described and the performance is measured. The default behavior provides excellent performance on a single disk – 50 MBps both reading and writing. Using large request sizes and doing file pre-allocation when possible have quantifiable benefits. When one considers disk arrays, .NET unbuffered IO delivers 800 MBps on a 16-disk array, but buffered IO delivers about 12% of that performance. Consequently, high-performance file and database utilities are still forced to use unbuffered IO for maximum sequential performance. The report is accompanied by downloadable source code that demonstrates the concepts and code that was used to obtain these measurements.

# Sequential File Programming Patterns and Performance with .NET

Peter Kukol, Jim Gray
Microsoft Research
{PeterKu, Gray} @Microsoft.com
December 2004

## 1. Introduction

Sequential file access is very common. Sequential file performance is critical for gigabyte-scale and terabyte-scale files; it can mean the difference between a task running in minutes or in days. This is the third in a series of articles that explores high-performance sequential file access on Windows™ file systems. The original paper, written in 1997 [Riedell97], studied Windows NT4 on a 200 MHz Pentium™ processor accessing "high-performance" 4 GB SCSI disks that delivered 7 MBps and cost more than $1,000 each. The Year 2000 study [Chung00] looked at Windows2000™ operating on dual 750 MHz processors accessing 27GB ATA disks that delivered 19 MBps and cost $400. This article examines WindowsXP™ and Windows Server 2003™ on dual 2.8 GHz processors accessing 250 GB SATA disks delivering 50 MBps and costing $130 each. Previous articles explained how to use low-level programming to trick the operating system into giving you good performance. The theme of this article is that the default behavior gives great performance, in large part because the hardware and software have evolved considerably over the years. So the article is really about how to write simple sequential file access programs on Windows™ systems using the .NET framework. It covers sequential text and binary access as well more advanced topics such as un-buffered access. It measures the speed and overhead impacts of block size, fragmentation, and other parameters. The concepts and techniques are illustrated using simplified C# code snippets available for download as a companion to this article [download].

## 2. Buffered File I/O

Sequential file access is very predictable, one can pre-fetch the next read and one can stream the sequence of writes. Randomly reading a disk, 8KB at a time, retrieves about one megabyte of data per second. Sequential access delivers 50 times more data per second. This sequential:random performance ratio is growing as technology improves disk densities and as disks spin faster. Applications are increasingly learning to buffer the "hot" data in main memory and sequentially pre-fetch data from and post-write data to disk.

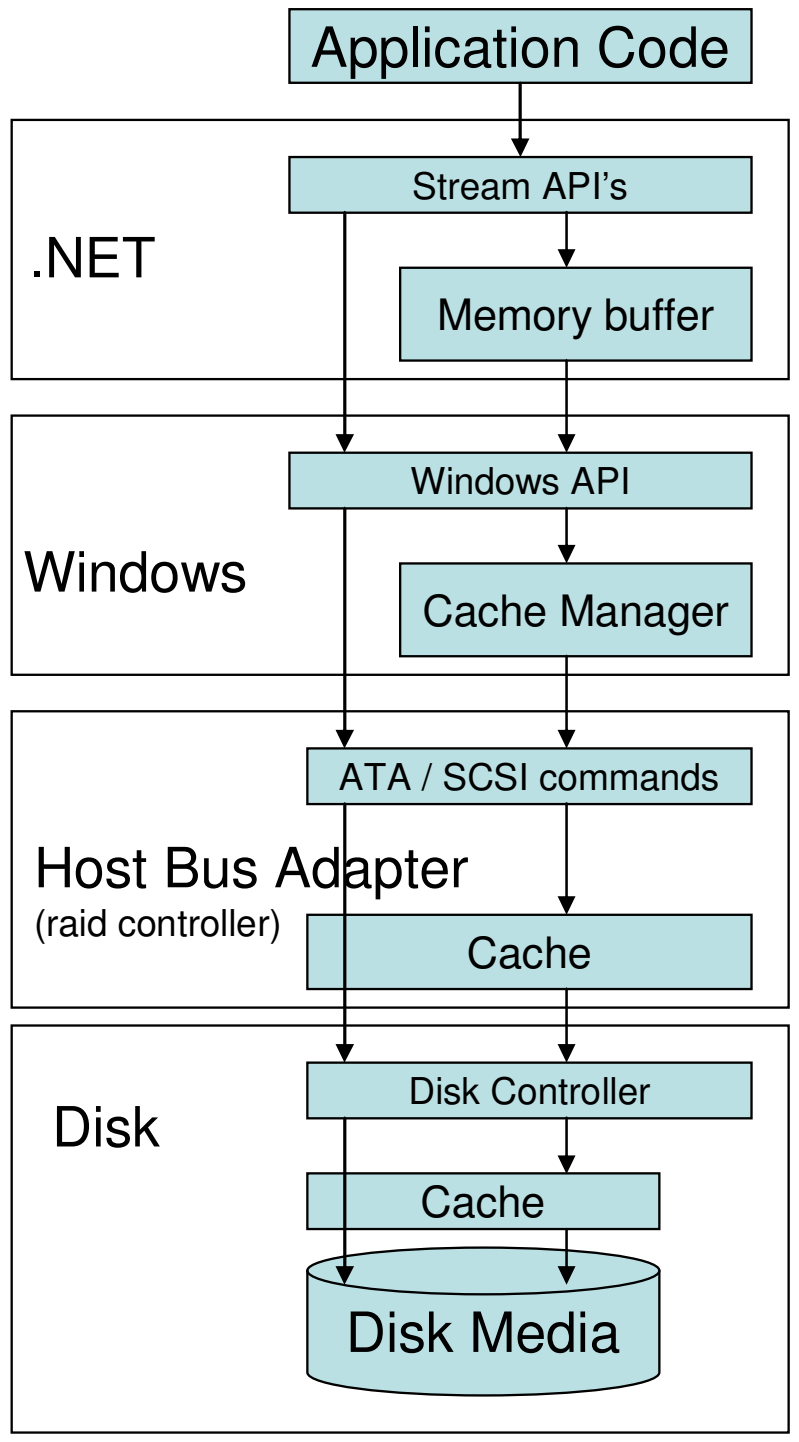

Figure 1: Hardware and software layers and caching in the disk IO path.

Like most runtimes, the .NET framework and Windows does this buffering for you when it detects a sequential file access pattern. As Figure 1 shows, the lower layers of the IO stack perform additional buffering. You might look at Figure 1 and say: "All those layers mean bad performance." Certainly, that is what our intuition tells us. But surprisingly most of the layers "get out of the way" in the common path, so the actual cost-per-byte is very low for sequential IO; yet, the layers provide excellent default behavior.

The main effect of buffering is to combine small logical read and write requests into fewer-larger physical disk I/O requests. This avoids reading the disk when the data is already in memory, thus improving performance. As an extreme example, consider a file being written one-byte-at-a-time. Without buffering, every write request would read a block from the disk, modify a byte, and then write the block back to the disk. Buffering combines thousands of such reads and writes into a single write that just replaces the block-values on disk (without ever having to read the old values of the blocks). The .NET runtime *stream* classes and Windows file system provide this buffering by default.

Buffering uses extra memory space, extra memory bandwidth, and extra CPU cycles. Seven years ago, this overhead was an important issue for most applications [Riedel97]. But, as explained in the first paragraph, processor speeds have improved 28-fold while disk speeds have improved a "mere" seven-fold. Measured in relative terms, disks have become four times slower than processors over the last decade – so sacrificing some processor and memory performance in exchange for better disk performance is a good bargain [Patterson]. It is RARE that a modern system is cpu-bound.

Our measurements and experience suggest that the cost of buffering is relatively minor and that the benefits almost always outweigh the costs. Thus, the default buffering should used unless measurements conclusively prove that its performance is significantly worse – a rare event. If your program is waiting, it is likely waiting for network or disk activity rather than waiting for a cpu. There are scenarios, notably in server-oriented transaction processing systems, where disabling buffering is appropriate. Sections 7and 8 quantify buffering costs so that you can evaluate this tradeoff for your application.

## 3. Sequentially Reading a Binary File

Opening a binary file and creating a stream to read its contents can be done in one step by creating a new instance of the FileStream class. The FileStream() constructor has many flavors (overloaded versions); let's use the simplest one at first:

```
string          fileName = @"C:\TEMP\TEST.DAT";
...
FileStream fs = new FileStream(fileName,           // name of file
                               FileMode.Open);      // mode (open/create/etc)
```

The only required arguments are the file name and the open 'mode'. The file name is a string for the full path to the file or it is interpreted relative to the current directory search path. The string constant used above is preceded by "@" to avoid needing double back-slashes in the file name like this: `"C:\\temp\\test.dat"` (the "@" notation is unique to C#). The file name is usually a path on a local disk, but it may be on a network share (e.g. `@"\\server\share\test.dat"`). In Windows, file names are not case sensitive.

The second parameter is a 'FileMode' enumeration value. The most common file modes are:

| | |
|---|---|
| **Open** | The file must already exist. Used to access existing files. |
| **Create** | If the file already exists truncate it, otherwise create it. (It is like CreateNew or Truncate.) |
| **CreateNew** | A new file will be created. An exception is thrown if the file already exists. Avoids over-writing existing files. |
| **OpenOrCreate** | Open an existing file; if it does not exist create an empty file. (It is like CreateNew or Append.) |
| **Append** | If the file exists, it is opened and data will be appended at its end.<br>If the file doesn't already exist, a new one is created. (It is like OpenOrCreate, but writes at the end.) |
| **Truncate** | The file must already exist. Open and truncate the current file contents. |

Opening a file may fail for several reasons. The file may not exist or the path may not be valid or you may not be authorized, or... Thus, the code should be wrapped in an exception handler. Ideally the handler would deal with each specific exception, but a simple handler that catches all exceptions (and displays the exception string before exiting) is the minimum requirement:

```
try { FileStream fs = new FileStream(fileName, FileMode.Open); }
catch (Exception e) {
        Console.WriteLine("Error opening file '{0}'. \n {1}",     fileName, e);
        throw new FileNotFoundException("Error opening file: " + fileName, e );
}
```

Once the FileStream is open, the basic choices are to read byte-at-a-time, line-at-a-time (if it is text), or byte-array–at-a-time. The easiest approach is to read one byte at a time:

```
int nextByte = 0;                              // holds next byte in the stream (or -1)
while (0 <= (nextByte = fs.ReadByte())) {      // ReadByte()returns -1 at end of file
    /* ... process 'nextByte' ... */
}
```

There is substantial overhead associated with reading each byte individually (see Figure 3 in section 8). If this overhead is an issue, an alternative is to read line-at-a-time or an entire byte array each time (and process each byte with an inner loop). The following two snippets demonstrate those metaphors:

```
using (StreamReader sr = new StreamReader(fileName)){
        string line;
        while ((line = sr.ReadLine()) != null) {
        /* ... process bytes line[0]… line[line.Length-1] */
        }       }
```

and:

```
int fileIOblockSize = 64 * 1024;                   // read up to 64KB each time
byte [] IObuff = new byte[fileIOblockSize];        // buffer to hold bytes
while (true) {
    int readCount = fs.Read(IObuff, 0, IObuff.Length);
    if (readCount < 0)
          break;
    /* ... process ' IObuff[i]' for i = 0...readCount-1*/
    }
```

The FileStream() constructor has several overloaded versions that let you control how the file is accessed. For example, to sequentially read a large file we can create the stream as follows:

```
FileStream fs = new FileStream(fileName,              // name of file
                        FileMode.Open,                // open existing file
                        FileAccess.Read,              // read-only access
                        FileShare.None,               // no sharing
                        2 << 18,                      // block transfer size = 256 KB
                        FileOptions.SequentialScan); // sequential access
```

This explicitly sets the IO transfers to be 256 KB. A rule of thumb is that larger transfer sizes are better (within reason), and powers of 2 work best. The measurements in Section 8 are a guide to selecting a good transfer size. The `SequentialScan` flag hints that access to the file will be sequential (recommending to the file system that it pre-fetch and post-write the data in large transfers).

## 4. Creating and sequentially writing a binary file

Sequential file writing is very similar to reading, except that proper buffering is extremely important. Without buffering, when a program changes just one byte, the file system must fetch the disk block that contains the byte, modify the block and then rewrite it. Buffering avoids this read-modify-write IO behavior. Windows and .NET provide write buffering by default, so that whole blocks are written and so that there are no extra reads if the block is being replaced. Assuming the FileStream `fs` has been opened, the byte-at-a-time write code is:

```
Random rand = new Random();                        // seed a random number generator
for (int j = 0; j < 100; j++) {                    // write 100 random bytes
      byte nextByte = (byte)(random.Next() % 256); // generate a random byte
      fs.WriteByte(nextByte);                      // write it
}
```

It is often better to accumulate an entire buffer of data, and write the entire buffer out at once:

```
byte[] IObuff = new byte[fileIOblockSize];            // buffer holds many bytes
for (int writeCount= 0; writeCount < fileIOblockSize*10; // write 10 buffers
                   writeCount += fileIOblockSize) {
    for (int k = 0; k < IObuff.Length; k++)           // fill the "write" buffer
        IObuff[k] = (byte)(random.Next() % 256);      // with random bytes
    fs.Write(IObuff, 0, IObuff.Length);               // write out entire buffer
}
```

New FileStream data isn't immediately reflected on the physical disk media. The new data is buffered in main memory. The .NET framework will *flush* the buffer under various conditions. Modified data are written out to the physical media by *lazy* background threads in the framework and the file system – a two-stage buffering process. It may be many seconds before modified data is written to disk; indeed, a temporary file may never be written if it is immediately deleted. The `Flush()` stream method returns when all that stream's .NET buffers have been written to the file system. To force the file system write to disk controller, you must call the `FlushFileBuffers()` Windows API (see the `IOspeed.fileExtend()` program for an example of this [download]). It is likely that `Flush()` will be overloaded to have a full-flush option in the future. Forcing the disk controller to write to disk is problematic – some disks and controllers observe the "force unit access" option in SCSI and SATA, but some do not. The NTFS flush() generates this command but it is often ignored by the hardware.

## 5. Reading and writing typed binary data

When integers, floats, or other values are written to or read from files, one option is to convert the values to text strings and use FileStreams. The problem with doing that is that you have to *serialize* the data yourself, converting between the value and its representation as a series of bytes in the `byte[]` buffer. This can be tedious and error-prone. Fortunately the framework provides a convenient pair of classes, `BinaryReader()` and `BinaryWriter()`, that read and write binary data. You simply "wrap" an instance of the binary class around our FileStream and then you can directly read and write any built-in type, including integers, floats, decimals, datetime, and strings.

The `FileStream.Write()` method can be called with an argument of any base type. The method's properly overloaded version will output the binary value. Reading binary values from a stream is the mirror image of this; calling the appropriate reader function such as `ReadInt16()` or `ReadString()` on a `BinaryReader` as shown in the following examples.

Writing typed data to a file.

```
// Open a file for writing
FileStream fs = new FileStream("...",
                         FileMode.CreateNew,
                         FileAccess.Write);
// Create a binary writer on the file stream
BinaryWriter bw = new BinaryWriter(fs);
// Write an integer value to the stream
uint integer_val = <whatever>;
bw.Write(integer_val);
// Write a string value to the stream
string string_val = <whatever>;
bw.Write(string_val);
… etc …
bw.Close();     // Close the stream
```

Reading the data from the file.

```
// Open a file for reading
FileStream fs = new FileStream("...",
                         FileMode.Open,
                         FileAccess.Read);
// Create a binary writer on the file stream
BinaryReader br = new BinaryReader(fs);
// Read an integer value to the stream
uint integer_val =  br.ReadUInt32();
// Read a string value to the stream
string string_val = br.ReadString();
… etc …
// Close the binary reader
br.Close();
```

## 6. Reading and writing text data

Text files are sequences of lines, strings terminated by a <new-line, carriage-return> character pair (UNIX lines are terminated by just a new-line character.) Text files can be read as a FileStream as described above; but, the `StreamReader` and `StreamWriter` classes also implement simple text-oriented file I/O and handle things such as file encoding (ASCII vs. Unicode and so on). Once you've opened a file for input or output and have a FileStream instance for it, you can wrap it in a StreamReader or StreamWriter instance and then easily read or write text lines. Here is an example that counts the lines in a file:

```
FileStream   fs = new FileStream(fileName, FileMode.Open);
StreamReader sr = new StreamReader(fs);
ulong        lc = 0;
while (sr.ReadLine() != null)
    lc++;
Console.WriteLine("The file '" + fileName + "' contains " + lc + " lines of text");
```

## 7. Summary of Simple Sequential File Access Programs

The previous sections showed the rudiments of creating, writing, and reading files in .NET. The next section presents measurements showing that this simple approach delivers impressive performance. It is rare that an application needs more than this direct approach. But, section 9 presents advanced topics that may be useful to very data intensive applications. To summarize, the following simple program shows how to sequentially create and read a file. The error handling has been removed to simplify the presentation.

```
using System;
using System.IO;

class Examples {
   static void Main(string[] args) {
      string filename = @"C:\TEMP\TEST.DAT";   // a file name
      FileStream fs = new FileStream(filename,// name of file
                            FileMode.Create);// mode (create a new file)
      for (int i = 0; i < 100; i++)            // write 100 bytes in the file
         fs.WriteByte((byte)'a');              // each byte is an "a"
      fs.Position = 0;                         // reposition at start of file.
      while(fs.Position<fs.Length)             // read to end of file
         {int nextByte = fs.ReadByte();}       // one byte at a time
      fs.Close();                              // close the file
      File.Delete(filename);                   // delete the file we just created
   }
}
```

## 8. Performance measurements

The performance of the simple file access programs was measured using Beta 1 of the .NET Framework Version 2.0 under Windows XP SP2 on the following hardware:

Tyan S2882 motherboard
Dual AMD Opteron 246 CPU's
2 GB PC3200 DDR RAM
SuperMicro MV-SATA disk controller
Maxtor 250 GB 7200 RPM SATA disk

The write benchmarks were run on a freshly formatted 250GB volume.  The 30GB test file was recreated every time unless otherwise noted.  The file sizes are large enough to overflow any caches  Section 10 discusses the effects of disk fragmentation.

The performance results used the programs at the download site and the detailed spreadsheet of measurements is also on that site [download].

Figure 2 shows the sequential FileStream speed in MB/sec vs. buffer size; the buffer size varied from 1 byte through 4 MB. Note that the first buffer size entry (labeled '1B') corresponds to the `readByte()` or `writeByte()` case.  All the other entries correspond to reading or writing a `byte[]` buffer of the indicated size.  The graph shows that, except for very small and very large buffers, the speed is nearly constant at about 50 MB/sec.

The slightly higher write speed is due to Windows and disk controller caching.  When a write is posted control usually returns immediately to the program which can then continue; whereas, when a synchronous read is issued the program has to wait until all of the data has arrived.

Figure 3 shows the average cost, in CPU cycles per byte of FileStream sequential reads and writes. It shows that using larger buffers reduces the per-byte overhead significantly.  Specifically, the overhead stabilizes at around 5 cycles per byte for writes and 10 cycles per byte for reads at request sizes of 64 KB.  This suggests that a dual 2 GHz cpu like the one benchmarked here could sustain over 1GBps of file activity. Other measurements have shown this to be the case. [Kukol04].  Most applications run at 1% or 10% of that speed, so it seems FileStream access is adequate for almost any application.  As shown in Figures 6 and 7, a file striped across 16 disk drives delivers 800 MBps and uses about 30% of a processor – when those experiments are done with buffered IO the speed is dramatically less – about 100 MBps vs 800 MBps – so for now, the .NET runtime is OK for single disks, but un-buffered IO is needed to drive disk arrays at speed..

To summarize: for simple sequential file access it's a good idea to use the FileStream class with a request size of 1 KB or larger. Request sizes of 64KB or larger have minimal cpu overhead.

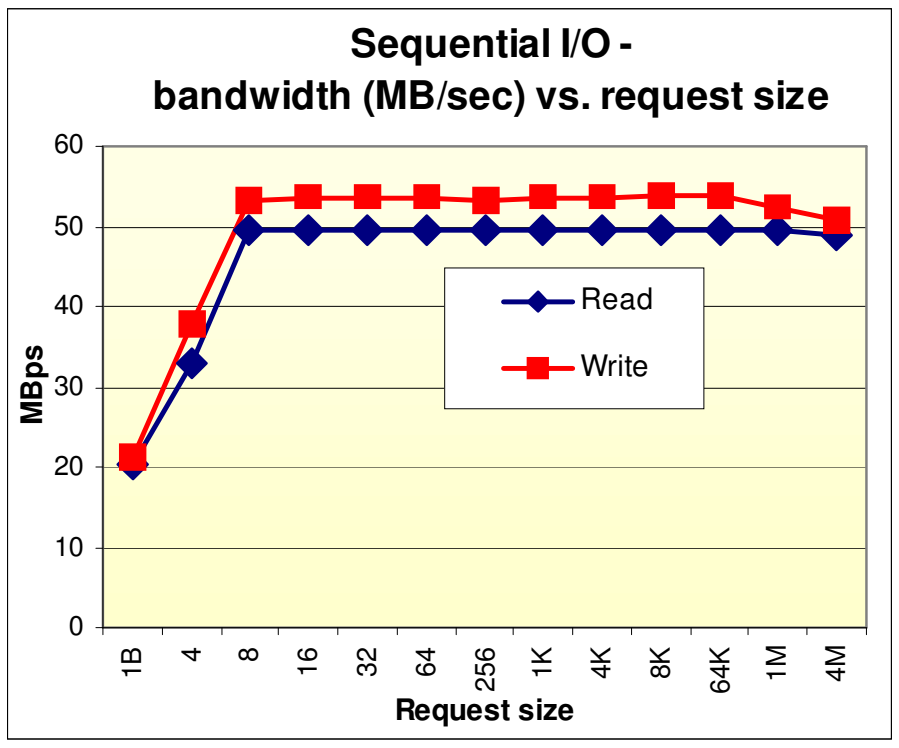

Figure 2: Buffered FileStream bandwidth vs. request size. Beyond 8-byte requests, the system runs at disk speed of around 50 MBps.

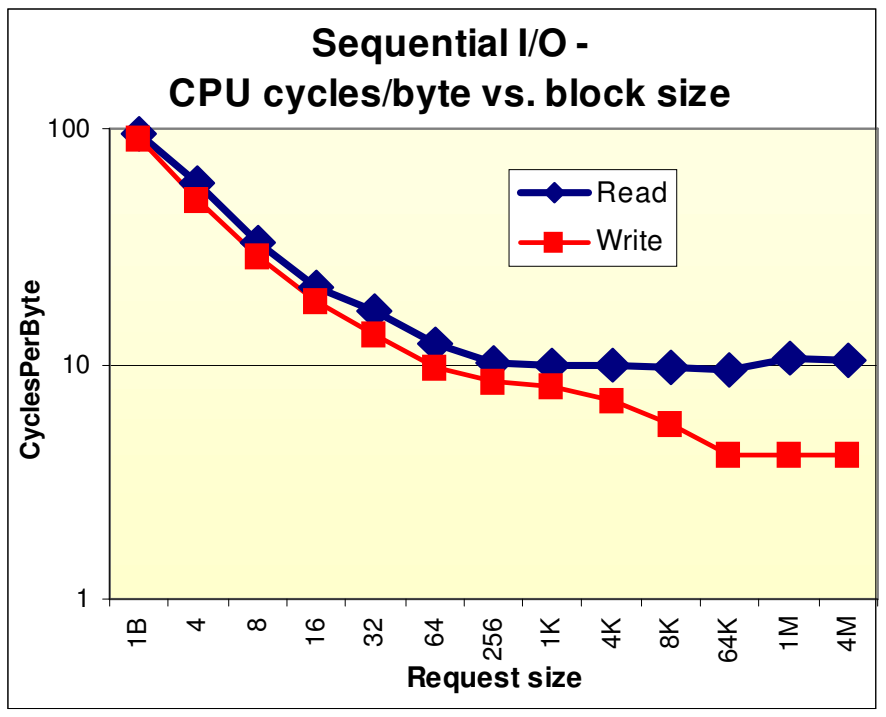

Figure 3: CPU consumption per byte of IO vs. request size. At 4KB and larger buffers the overhead stabilizes at about 5 or 10 clocks per byte.

# 9. Un-buffered file performance measurements

FileStream buffering can be disabled for that stream when it is opened. This bypasses the file cache and avoids the consequent overhead of moving data between the NTFS cache, the .NET cache and the application data space. In un-buffered IO, the disk data moves directly between the application's address space and the device (the device adapter in Figure 1) without any intermediate copying. To repeat the conclusion of the last section, the FileStream class does a fine job. Most applications do not need or want un-buffered IO. But, some applications like database systems and file copy utilities want the performance and control un-buffered IO offers.

There is no simple way to disable FileStream buffering in the V2 .NET framework. One must invoke the Windows file system directly to obtain an un-buffered file handle and then 'wrap' the result in a FileStream as follows:

```
[DllImport("kernel32", SetLastError=true)]
static extern unsafe SafeFileHandle CreateFile(
        string FileName,            // file name
        uint DesiredAccess,         // access mode
        uint ShareMode,             // share mode
        IntPtr SecurityAttributes,  // Security Attr
        uint CreationDisposition,   // how to create
        uint FlagsAndAttributes,    // file attributes
        SafeFileHandle hTemplate    // template file
        );

SafeFileHandle handle = CreateFile(FileName,
                               FileAccess.Read,
                               FileShare.None,
                               IntPtr.Zero,
                               FileMode.Open,
                                FILE_FLAG_NO_BUFFERING,
                               null);
FileStream stream = new FileStream(handle,
                               FileAccess.Read,
                               true,
                               4096);
```

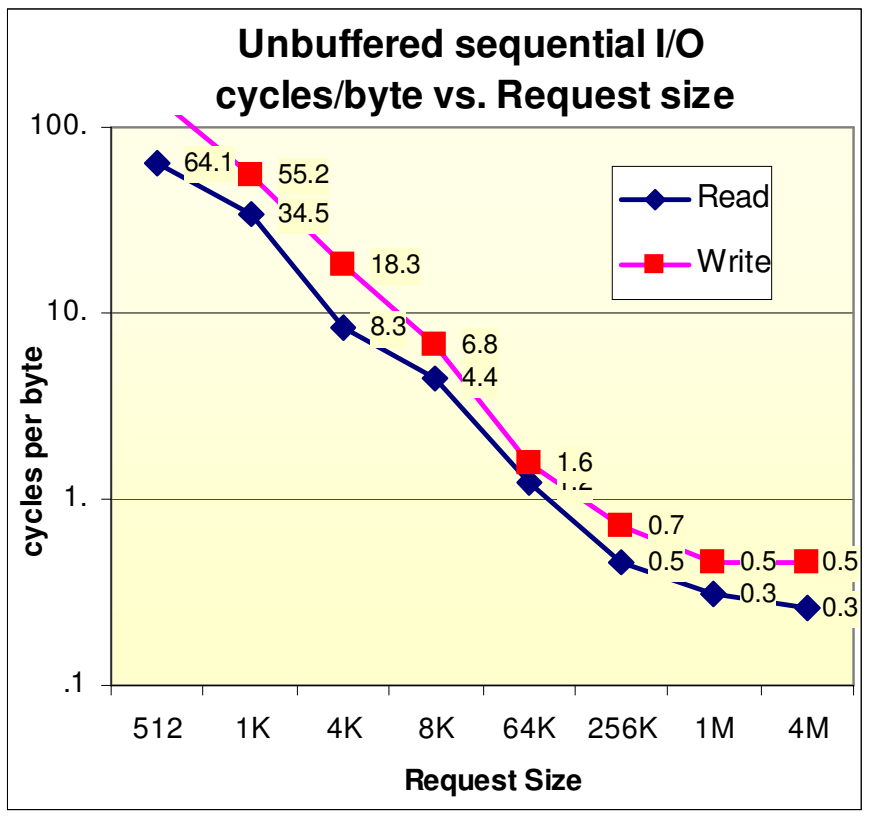

Figure 4: Un-buffered IO CPU cost is 10x less than buffered IO for large requests.

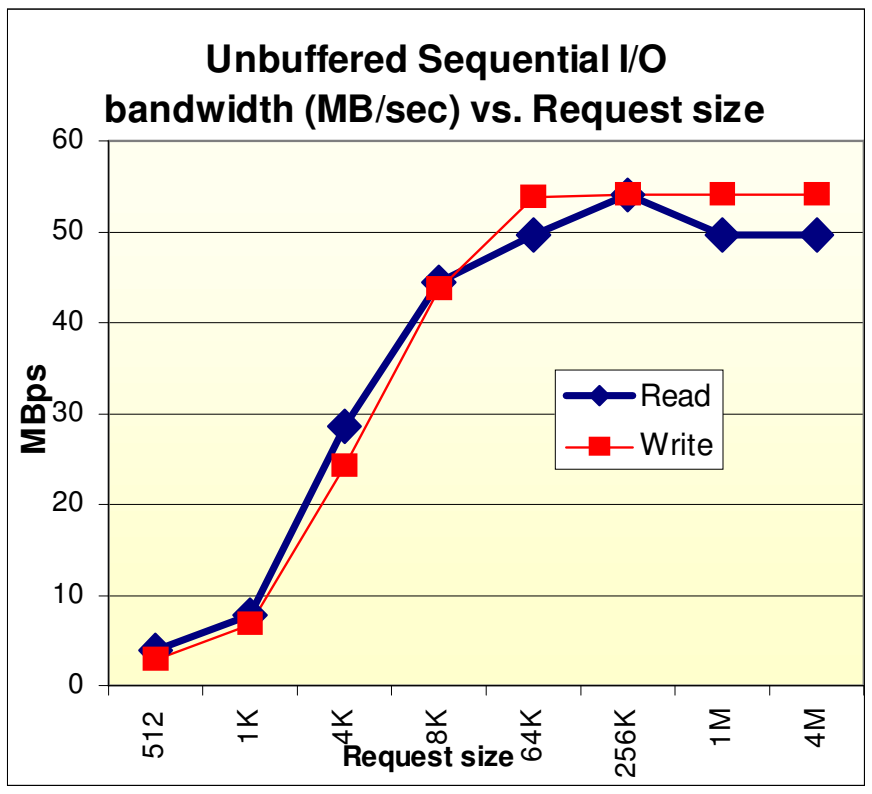

Figure 5: Un-buffered IO has good throughput at 64KB request sizes.

Calling `CreateFile()` with the `FILE_FLAG_NO_BUFFERING` flag tells the file system to bypass all software memory caching for the file. The 'true' value passed as the third argument to the `FileStream` constructor indicates that the stream should take ownership of the file handle, meaning that the file handle will automatically be closed when the stream is closed. After this hocus-pocus, the un-buffered file stream is read and written in the same way as any other.

Un-buffered I/O goes almost directly to the hardware, so the request buffers must be aligned to a sector boundary (both in memory and within the file), and the request size must be a multiple of the volume's sector size. `VirutalAlloc()` returns storage aligned to a page boundary – so unmanaged request buffers can be allocated from such page aligned storage. Today sectors on most disks are 512 bytes, but in the future they may well be much larger so a 64KB alignment is recommended. `IOspeed.DriveSectSize()` determines the device sector size of a file [download].

Figure 4 shows the per-byte overhead of reading/writing files with buffering disabled. The overhead of un-buffered IO using large requests is approximately 0.2 – 0.3 cycles per byte.

Not surprisingly, Figure 5 shows that a 64KB transfer size is necessary to achieve the highest speed with un-buffered I/O. A rule of thumb is that the minimum recommended transfer size is 64 KB and that bigger transfers are generally better.

Un-buffered file streams can drive very fast disk configurations. Using Windows disk manager, we created a striped NTFS volume (often called "software RAID 0") that spanned 16 physical drives spread across two 8-port SuperMicro SATA controllers in two PCI-X busses and measured file stream read and write speeds. Figures 6 and 7 show that the file can be read at about 800 MB/sec and written at about 400 MB/sec using un-buffered I/O – the limited write speed is an artifact of the disk drives we used (8 of them were Hitachi 400GB drives that write at 25MBps and so slow the entire array to that write speed.) When large requests are used, the CPU overhead (clocks per byte) is quite low – about 0.7 clocks per byte. Striped NTFS volumes require the operating system do more work to distribute the I/O requests across the physical media and this is reflected in a slightly higher overhead when Figure 7 is compared to Figure 4.

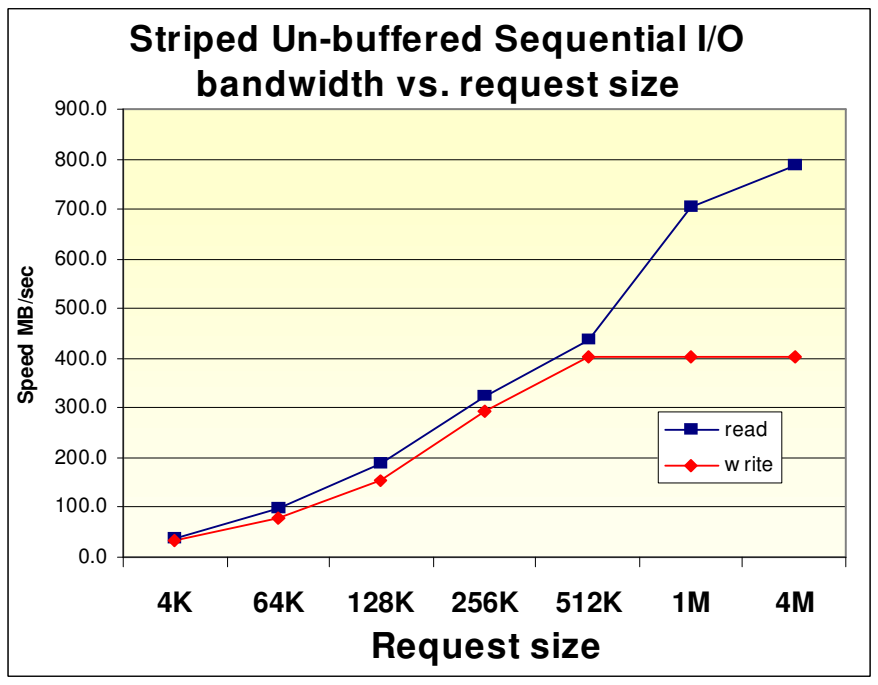

Figure 6: Striped volume throughput. The 400 MBps write plateau reflects the speed disk speed limit of eight of the 16 drives.

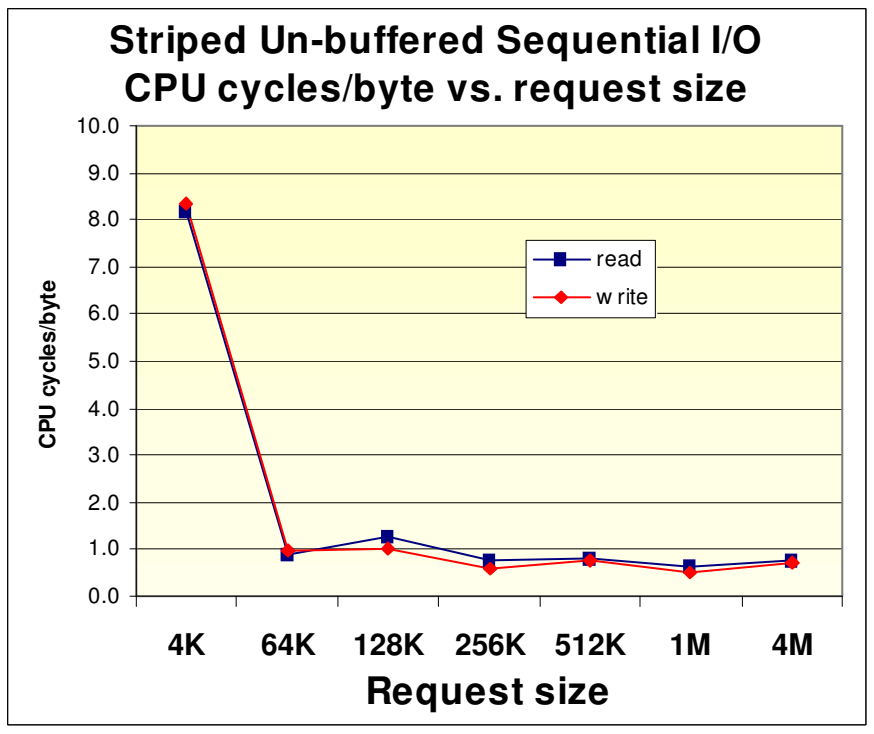

Figure 7: Striped volume cycles per byte.

These experiments indicate that a simple C# file stream program is capable of reading a large disk array at speeds on the order of 800 MB/sec with low processor overhead. They also show that the controller and bus bottlenecks observed 5 years ago by Chung [Chung00] have been addressed. Most controllers can support 8 drives and most 64-bit PCI-X can support 750MBps transfers. Figure 6 requires two such busses, but with PCI-Express will easily support the bandwidth of 8 future disks.

## 10. The cost of file fragmentation

When a large file is created by incrementally appending data – new file *extents* are allocated as the file grows. These extents are not necessarily contiguous – they are allocated using a best-fit algorithm. If a file has many non-contiguous extents, we say the file is *fragmented*. A sequential scan of a fragmented file can be a random scan of the disk as the scan reads each fragment in turn. Fragmentation can cause a significant performance loss both when the file is initially created and when it is later accessed.

Simple techniques can reduce fragmentation when creating or extending large files, and utilities can reduce fragmentation of existing files by reorganizing the disk. When the approximate file size of a file is known in advance it is best to tell the file system the estimated size as soon as possible (ideally, immediately after the file is opened.) This lets the file system efficiently pre-allocate the largest possible chunks (the fewest fragments) of physical media to hold all of the file contents and thereby reduce fragmentation. The simplest way to do this is to extend the file to the known final size. With a FileStream, this is done by invoking the `SetLength()` method. The following code creates and allocates a 128 megabyte file:

```
FileStream fs = new FileStream(fileName, FileMode.OpenOrCreate);
fs.SetLength(128 * 1024 * 1024);
```

Figure 8 shows how file creation speed improves when the final file size is set at creation time. The tests were run on a freshly formatted volume ("clean disk") as well as a volume that was fragmented with a tool we built [download]. Since fragmentation in the real world is a stochastic process, the tool repeatedly creates and deletes files randomly. We then ran the tests (taking the median of all the measurements.) As can be seen from the graphs, one should make at least 1 KB write requests to get decent performance. Beyond that, if the disk is fragmented, it pays to tell the file system about the final file. Declaring the file size at creation time reduces the disk fragmentation slow-down from about 25% to about 15% -- (46MBps and 39MBps respectively). The 15% pre-allocated file slowdown is caused by disk zoning – the innermost zone runs at 34 MBps while the outermost zone runs at 57MBps – the average speed of 46MBps is almost exactly the "red line" speed measured for preallocated files in Figure 8. On a "clean disk," files are allocated in the outer band of the disk. On a fragmented disk, files are typically allocated in the middle and inner disk zones that have lower byte transfer rates. We conclude from this that fragmentation reduces sequential performance by about 13%.

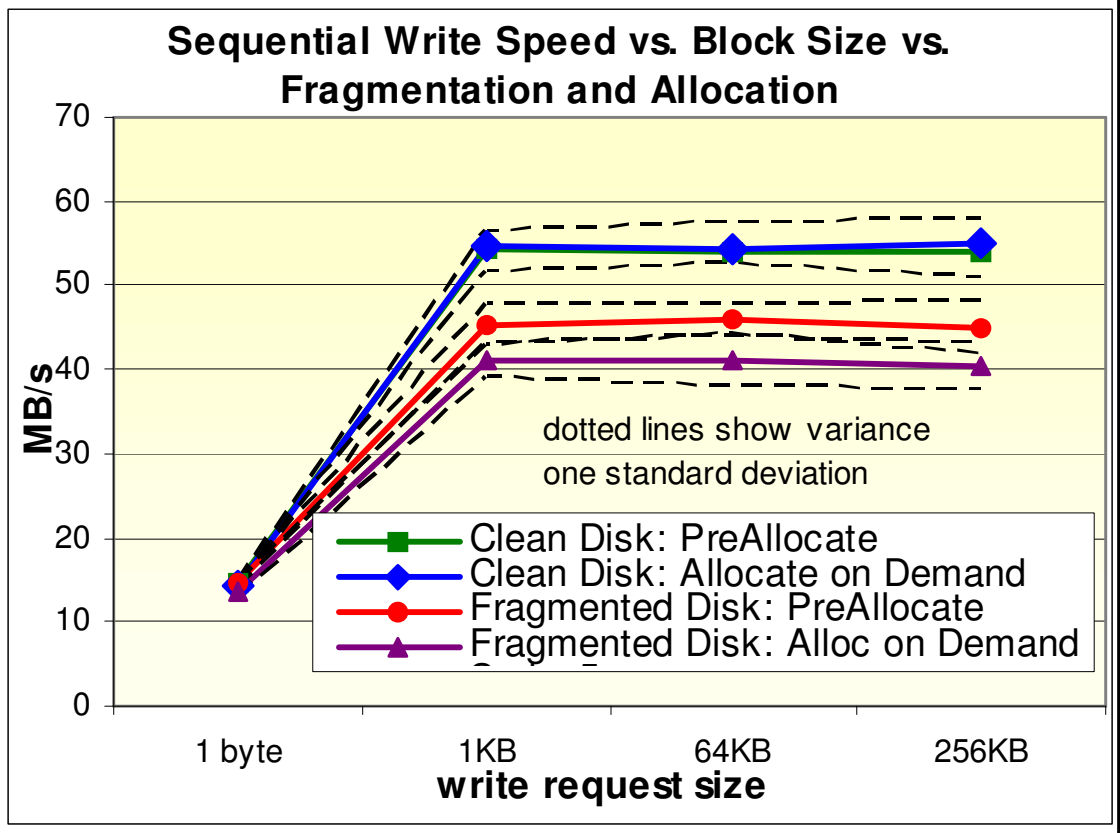

Figure 8: File creation speed vs. write-request size, disk fragmentation and pre-allocation of file length. The dotted lines show the standard deviation of each measurement.

# 11. Summary

Seven years ago, one had to be a guru to read or write disks at 7MBps [Riedel97]. That programming style is still possible (see the Appendix for an example.) Now, the out-of-the-box default performance is 7x that and disk arrays deliver over 1 GBps. In part this is because processors, disks, and controllers have improved enormously; but, in large part it reflects the fact that the software stack described in Figure 1 has been streamlined and redesigned so that it usually does the right thing. This evolution is not complete (e.g. the flush option does not flush the NTFS cache), but those details are being repaired as we write this article.

In 1997 and 2000, 4 disks could saturate a disk controller and one controller could saturate a bus. Modern controllers can handle the bandwidth of 8 disks and modern busses can handle the bandwidth of two controllers.

To summarize our findings:

(1) For single disks, use the defaults of the .NET framework – they deliver excellent performance for sequential file access.
(2) Pre-allocate large sequential files (using the `SetLength()` method) when the file is created. This typically improves speed by about 13% when compared to a fragmented file.
(3) At least for now, disk arrays require un-buffered IO to achieve the highest performance. Buffered IO can be 8x slower than un-buffered IO. We expect this problem will be addressed in later releases of the .NET framework.
(4) If you do your own buffering, use large request sizes (64KB is a good place to start).
(5) Using the .NET framework, a single processor can read and write a disk array at over 800 MBps using un-buffered IO.

One can characterize the three articles in this series as:

Riedel 1997: Windows can drive SCSI disks and disk arrays as fast as they can go, if you are a guru. Controllers and busses are bottlenecks.
Chung 2000: IDE disks are slower than SCSI but they have great price-performance. Controllers and busses are bottlenecks.
Kukol 2004: .NET runtime delivers excellent performance by default and it is easy to configure balanced IO systems with commodity SATA disks and controllers.

Certainly, there will be an opportunity to redo these experiments in 2010, but we are unclear what the future holds. We expect the .NET framework to fix the buffered disk array performance issues and to make it easier to disable all buffering. Since disk capacities will likely be in the 3 terabyte range by then, we assume issues of file placement, archiving, and snapshoting will be the dominant concerns.

# References

[Chung00] L. Chung, J. Gray, B. Worthington, R. Horst, Windows 2000 Disk IO Performance, MSR-TR-2000-55, September2000

[Download] http://research.microsoft.com/research/downloads/ and http://research.microsoft.com/research/downloads/download.aspx?FUID={A6F86A1E-0278-4C72-9300-F747251C3BF0}

[Kukol04] P. Kukol, J. Gray, "Sequential Disk IO Tests for GBps Land Speed Record," March 2004, MSR-TR-2004-62 http://research.microsoft.com/research/pubs/view.aspx?type=Technical%20Report&id=766

[Patterson] D. Patterson, "Latency Lags Bandwidth," ROC retreat, January 2004, http://roc.cs.berkeley.edu/retreats/winter_04/posters/pattrsn_BWv1b.doc

[Riedel97] E. Reidel, C. VanIngen, J. Gray, A Performance Study of Sequential IO on WindowsNT™ 4.0, MSR-TR-97-34, September 1997

# Acknowledgments

Leonard Chung, Catharine vanIngen, and Bruce Worthington were very helpful in designing these experiments and made several consturcive suggestions for improving the presentation.

# Appendix

The following code shows what is needed to do asynchronous and un-buffered IO. It gives the example of a program that copies one file to another. The reader launches N reads and then, as they complete, it writes the buffer to the target file. The asynchronous write completion callback issues the next read on that buffer. This code has no error handling; it is literally the shortest program we could write (error handling is included in the download version of this program [download].) **It is included here to persuade you that you do not want to do this unless you really are in pain.** Buffering allows you to overlap reads and writes and get most of the benefits of this asynchronous code.

```
using System;
using System.IO;
using System.Threading;

namespace AsyncIO {
    class FileCopy {
    // globals
        const int          BUFFERS = 4;           // number of outstanding requests
        const int          BUFFER_SIZE = 1<<20;  // request size, one megabyte
        public static FileStream source;         // source file stream
        public static FileStream target;         // target file stream
        public static long totalBytes   = 0;     // total bytes to process
        public static long bytesRead    = 0;     // bytes read so far
        public static long bytesWritten = 0;     // bytes written so far
        public static Object WriteCountMutex = new Object[0]; // mutex to protect count
        // Array of buffers and async results.
        public static AsyncRequestState [] request = new AsyncRequestState [BUFFERS];
        // structure to hold IO request buffer and result.
        public class AsyncRequestState {         // data that tracks each async request
            public byte[]          Buffer;       // IO buffer to hold read/write data
            public AutoResetEvent ReadLaunched; // Event signals start of read
            public long            bufferOffset; // buffer strides thru file BUFFERS*BUFFER_SIZE
            public IAsyncResult    ReadAsyncResult;// handle for read requests to EndRead() on.
            public AsyncRequestState(int i){     // constructor
                bufferOffset = i * BUFFER_SIZE; // offset in file where buffer reads/writes
                ReadLaunched = new AutoResetEvent(false); // semaphore says reading (not writing)
                Buffer = new byte[BUFFER_SIZE]; // allocates the buffer
        }   }                                    // end AsyncRequestState declaration
    // Asynchronous Callback completes writes and issues next read
        public static void WriteCompleteCallback(IAsyncResult ar) {
            lock (WriteCountMutex) {            // protect the shared variables
                int i = Convert.ToInt32(ar.AsyncState);    // get request index
                target.EndWrite(ar);            // mark the write complete
                bytesWritten += BUFFER_SIZE;   // advance bytes written
                request[i].bufferOffset += BUFFERS * BUFFER_SIZE; // stride to next slot
                if (request[i].bufferOffset < totalBytes) {       // if not all read, issue next read
                    source.Position=request[i].bufferOffset;      // issue read at that offset
                    request[i].ReadAsyncResult = source.BeginRead(request[i].Buffer,0, BUFFER_SIZE,null,i);
                    request[i].ReadLaunched.Set();
        }   }   }
    // main routine implements asynchronous File.Copy(@"C:\temp\source.dat", @"C:\temp\target.dat");
        static void Main(string[] args) {
            source = new FileStream(@"C:\source.dat",  // open source file
                            FileMode.Open,            // for read
                            FileAccess.Read,          //
                            FileShare.Read,           // allow other readers
                            BUFFER_SIZE,              // buffer size
                            true);                    // use async
            target = new FileStream(@"C:\target.dat",  // create target file
                            FileMode.CreateNew,       // fault if it exists
                            FileAccess.Write,         // will write the file
                            FileShare.None,           // exclusive access
                            BUFFER_SIZE,              // buffer size
                            true);                    // use async
            totalBytes = source.Length;               // Size of source file
            AsyncCallback writeCompleteCallback = new AsyncCallback(WriteCompleteCallback);
            for (int i = 0; i < BUFFERS ; i++)   request[i] = new AsyncRequestState(i);
            // launch initial async reads
            for (int i = 0; i < BUFFERS; i++) {        // no callback on reads.
                request[i].ReadAsyncResult = source.BeginRead(request[i].Buffer, 0, BUFFER_SIZE, null, i);
                request[i].ReadLaunched.Set();         // say that read is launched
            }
            // wait for the reads to complete in order, process buffer and then write it.
            for (int i = 0; (bytesRead < totalBytes ); i = (i+1) % BUFFERS) {
                request[i].ReadLaunched.WaitOne();     // wait for flag that says buffer is reading
                int bytes = source.EndRead(request[i].ReadAsyncResult); // wait for read complete
                bytesRead += bytes;                    // process the buffer <your code goes here>
                target.BeginWrite(request[i].Buffer, 0, bytes, writeCompleteCallback, i); // write it
            }                                          // end of reader loop
            while (pending > 0) Thread.Sleep(10);      // wait for all the writes to complete
            source.Close();  target.Close();           // close the files
        }                                              // end of async copy.
    }
}
```

# IOspeed.exe tests read and write bandwidth to a file

## Usage:

**`IOspeed [options] filePath`**

## Options:

```
-r[fileSize]  read optionally with file size to be created if needed (default=1G)
-w[fileSize]  write (default is read) optionally with file size to be written (default=1G)

-t<seconds>   test duration (default=30 seconds)
-b<size>      I/O block size (default=64K)
-a[count]     use asynch (overlapped) I/O, (Default is sync, default async depth is 4).
-d            disable .NET and NTFS disk caching (memory buffering)
-s[seekDistancePercentage] random I/O (default is sequential)
              optionally average seek distance percentage of filesize, default=100.

-x< fileSize> create/extend/fill a file to given size
-p< fileSize> like -x but preallocate file before writing (reduces fragmentation).

-c            touch every byte in the source file (more memory and cpu load)
-q            quiet mode (as opposed to verbose), write data as comma separated list.

sizes suffixed with K, M, or G are are interpreted as Kilo, Mega, or Giga bytes.
```

## Examples:

| | |
|---|---|
| Iospeed –t30 –b64K –r1G –s0 a.dat | // the default settings, same as next line |
| Iospeed a.dat | // sequential read a.dat, 64KB requests, buffering for 30 seconds |
| Iospeed –t60 -p100M a.dat | // preallocates 100MB file a.dat, then reads it for 60 seconds |
| Iospeed -t30 -w100M -p a.dat | // preallocates 100MB file a.dat, then writes it for 30 seconds |
| Iospeed –a2 -b256K a.dat | // 2-deep async read of a.dat in 256KB requests |

## Description

IOspeed tests read and write bandwidth to a file.

The *–r* and *–w* options measure read and write speed respectively. They can be modified by the *–t<seconds>* option to request a test duration different from the 30 second default, or with a *–a<count>* option to request asynchronous IO with the specified count of outstanding IOs, or with the *–d* option to disable .NET and NTFS file caching, or with the *–b* option to specify an IO request size other than the 64KB default, and with a *–s<percentage>* option to request random seeks between IOs (*-s0*) is the sequential default and *–s100* causes the seek to go to a random place on the disk. The percentage is a distance control, not a fraction of sequential-random. The *–c* option asks the cpu to "touch" every byte rather than just discarding the bytes from main memory, making the cpu and memory bandwidth load more realistic.

The default file size is 1GB. The *fileSize* option can override this value. If the file does not exist it is created with this size. If the file already exists, and it is smaller than the requested size, it is extended to have this size. These file extension and the write tests overwrite the file.

The *–x* and *–p* options test the speed of file extension with and without pre-allocation. If there is a preexisting file with that name, it is deleted and a new empty file is created. *–x* incrementally grows the file by appending to the end (using synchronous writes of the given block size). The *–p* option first sets the file length, thereby allocating fewer fragments and then acts just like the *–x* option.

**FragDisk.exe** fragments a disk by creating a large number nearly filling the disk and then deleting some of the files at random to create “holes”.

## Usage

```
FragDisk [options] directoryPath
```

## Options:

```
-m<count>   set max. number of files to create              (default=100000)
-Fm<size>   set min. file size (in MB)                      (default=1)
-FM<size>   set max. file size (in MB)                      (default=256)
-c<count>   set number of files per cycle                   (default=1000)
-d<count>   set max. number of files per directory          (default=100)
-s<count>   set max. number of sub-directories per directory (default=10)
-n<count>   set max. create/delete cycles, 0 = no limit     (default=0)
-k<pctg>    set percentage (1-99) of files to keep          (default=5)
-f<pctg>    set desired percentage (1-99) to fill the volume (default=70)
-r<value>   set random seed                                 (default=137)
```

## Example:

```
FragDisk -f95 -k10 c:\temp
```

## Description:

This code creates many directories and files (of various sizes) and then deletes a random subset of them thereby creating a fragmented disk. The file sizes are randomly chosen between 1MB and 256MB. FragDisk recursivly buids a directory tree with enough sub-directories in it so that the leaves can accommodate enough files to fill the disk to the desired fullness. It then crates enough files of sizes chosen randomly between 1MB and 256MB (or sizes specified via the –F command-line options) to fill the disk to –f percentage. It then deletes a random subset of these files so that the disk shrinks ot –k percent full. It repeats this process for –n cycles.

# IOexamples.exe source code examples from the paper.

## Usage

IOexamples.exe [fileName [ recordCount] ]

## Options:

```
default file name is:            C:\IO_Examples_temp.txt
default record count is          1,000,000
```

## Example:

IOexamples C:\temp\test.dat 10000000        -- creates and tests a file of 100m 100-byte records

## Description:

This code demonstrates different styles of and performance of sequential file IO on Windows NTFS file systems using .NET IO classes. They are the examples for the MSR Technical Report by Kukol and Gray titled: "Sequential File Programming Patterns and Performance with .NET". These are typicaly IO programming patterns.

This times the following cases:

Read | Write

byte-at-a-time | line-at-a-time | 64KB block-at-a-time.

The file-write-speed test builds the sort benchmark file specified at http://research.microsoft.com/barc/SortBenchmark/

**AsyncCopy.exe** source code for async copy shown on page 9.

### Usage

AsyncCopy

### Options:

none

### Example:

AsyncCopy -- creates and copies a 1GB file

### Description:

This code is the full project (with error handling) of the async copy program in the appendix of the Micrsoft Technical Report by Kukol and Gray titled: Sequential File Programming Patterns and Performance with .NET. The code is downloadable from: http://research.microsoft.com/research/downloads/. The program creates the 1GB file `C:\temp\source.dat` and copies it to `C:\temp\target.dat` using double buffering (four outstanding IOs.) In the end it deletes the source and target files.